\documentclass{ws-ijmpd}
\usepackage[super,compress]{cite}

\usepackage{amsmath,amssymb,amsfonts,latexsym}
\usepackage{graphicx}
\usepackage{color}

\def\noi{\noindent}
\newcommand{\foom}[1]{\protect\footnotemark[#1]}

\def\cm{\hspace*{1cm}}

\def\Jl#1#2{{\it #1\/} {\bf #2}\ }

\def\ApJ#1 {\Jl{Astroph. J.}{#1}}
\def\CQG#1 {\Jl{Class. Quantum Grav.}{#1}}
\def\DAN#1 {\Jl{Dokl. AN SSSR}{#1}}
\def\GC#1 {\Jl{Grav. Cosmol.}{#1}}
\def\GRG#1 {\Jl{Gen. Rel. Grav.}{#1}}
\def\JETF#1 {\Jl{Zh. Eksp. Teor. Fiz.}{#1}}
\def\JETP#1 {\Jl{Sov. Phys. JETP}{#1}}
\def\JHEP#1 {\Jl{JHEP}{#1}}
\def\JMP#1 {\Jl{J. Math. Phys.}{#1}}
\def\NCB#1 {\Jl{Nuovo Cim. B}{#1}}
\def\NPB#1 {\Jl{Nucl. Phys. B}{#1}}
\def\NP#1 {\Jl{Nucl. Phys.}{#1}}
\def\PLA#1 {\Jl{Phys. Lett. A}{#1}}
\def\PLB#1 {\Jl{Phys. Lett. B}{#1}}
\def\PRD#1 {\Jl{Phys. Rev. D}{#1}}
\def\PRL#1 {\Jl{Phys. Rev. Lett.}{#1}}

\def\al{&}
\def\lal{&&{}}

\def\eqs{Eqs.\,}
\def\beq{\begin{equation}}
\def\eeq{\end{equation}}
\def\bear{\begin{eqnarray}}
\def\bearr{\begin{eqnarray} \lal}
\def\ear{\end{eqnarray}}
\def\earn{\nonumber \end{eqnarray}}

\def\nnv{\nonumber\\[5pt] {}}
\def\nnn{\nonumber\\ \lal }
\def\nnnv{\nonumber\\[5pt] \lal }

\def\yyy{\\[5pt] \lal }
\def\eql{\al =\al}

\def\dst{\displaystyle}
\def\tst{\textstyle}
\def\fracd#1#2{{\dst\frac{#1}{#2}}}
\def\fract#1#2{{\tst\frac{#1}{#2}}}
\def\Half{{\fracd{1}{2}}}
\def\half{{\fract{1}{2}}}

\def\e{{\,\rm e}}
\def\d{\partial}

\def\arg{\mathop{\rm arg}\nolimits}

\def\diag{\mathop{\rm diag}\nolimits}

\def\const{{\rm const}}
\def\eps{\varepsilon}

\def\then{\ \Rightarrow\ }

\newcommand{\vars}[1]{\left\{\begin{array}{ll}#1\end{array}\right.}

\def\rf{\eqref}

\def\MN{^{\mu\nu}}
\def\mN{_\mu^\nu}

\def\D{{\mathbb D}}
\def\M{{\mathbb M}}
\def\R{{\mathbb R}}
\def\tK{{\widetilde K}}
\def\tT{{\widetilde T}}

\def\kappa{\varkappa}

\def\cy{cylindrical}
\def\cyl{cylindrically symmetric}

\def\wh{wormhole}
\def\whs{wormholes}
\def\asflat{asymptotically flat}

\begin{document}

\markboth{K.A. Bronnikov, S.V. Bolokhov and M.V. Skvortsova}
{Cylindrical wormholes: a search for viable phantom-free models in GR}


\title{CYLINDRICAL WORMHOLES: A SEARCH FOR VIABLE PHANTOM-FREE 
	MODELS IN GR}

\author{K. A. BRONNIKOV\foom 1}

\address{VNIIMS, 	Ozyornaya 46, Moscow 119361, Russia;\\
	Institute of Gravitation and Cosmology, Peoples' Friendship University of Russia\\
	(RUDN University), ul. Miklukho-Maklaya 6, Moscow 117198, Russia;\\
	National Research Nuclear University ``MEPhI''
		(Moscow Engineering Physics Institute),\\
		Kashirskoe sh. 31, Moscow 115409, Russia\\
		kb20@yandex.ru		}
		
\author{S.V. BOLOKHOV}		

\address{Institute of Gravitation and Cosmology, Peoples' Friendship University of Russia\\
	(RUDN University), ul. Miklukho-Maklaya 6, Moscow 117198, Russia;\\
		boloh@rambler.ru        }
		
\author{M.V. SKVORTSOVA}	
\address{Peoples' Friendship University of Russia	(RUDN University),\\ 
		ul. Miklukho-Maklaya 6, Moscow 117198, Russia;\\
		milenas577@mail.ru        }

\maketitle


\begin{abstract}
  The well-known problem of wormholes in general relativity (GR) is the necessity 
  of exotic matter, violating the Weak Energy Condition (WEC), for their support.
  This problem looks easier if, instead of island-like configurations, one considers
  string-like ones, among them, cylindrically symmetric space-times with rotation. 
  However, for cylindrical wormhole solutions a problem is the lacking asymptotic 
  flatness, making it impossible to observe their entrances as local objects in our 
  Universe. It was suggested to solve this problem by joining a wormhole solution
  to flat asymptotic regions at some surfaces $\Sigma_-$ and $\Sigma_+$ 
  on differents sides of the throat. The configuration then consists of three regions, 
  the internal one containing a throat and two flat external ones. We discuss different 
  kinds of source matter suitable for describing the internal regions of such models 
  (scalar fields, isotropic and anisotropic fluids) and 
  present two examples where the internal matter itself and the surface matter
  on both junction surfaces $\Sigma_\pm$ respect the WEC. 
  In one of these models the internal source is a stiff perfect fluid whose pressure  
  is equal to its energy density, in the other it is a special kind of anisotropic 
  fluid. Both models are free from closed timelike curves.
  We thus obtain examples of regular twice asymptotically flat wormhole models 
  in GR without exotic matter and without causality violations. 
\end{abstract}


\ccode{PACS numbers: 04.20.-q, 04.20.Jb, 04.40.Nr, 04.20.Gz}


\section{Introduction}

  The possible existence of traversable Lorentzian wormholes in the Universe 
  is one of the most intriguing predictions of modern physical theories which 
  identify the gravitational field with the properties of space-time geometry.  
  Such objects, being a kind of tunnels in space-time, can make close to each 
  other otherwise distant regions of space as well as different epochs of the 
  Universe evolution, thus acting as time machines.\cite{thorne,vis-book}.
  Some researchers consider possible observable effects of  \whs\ if they 
  really exist in space  \cite{sha, accr, kir-sa1, we-lens}, see also references therein.

  Gravitation is in general an attractive force and tends to focus particle 
  trajectories, whereas in a \wh\ geometry some of the paths first approximate 
  but then move away from each other. Such a behavior is a qualitative reason for 
  the necessity of unusual forms of matter, producing repulsion instead of
  attraction, to support the \wh\ geometry. In the framework of general relativity (GR),
  the necessity of matter violating the weak and null energy conditions 
  (WEC and NEC), called ``exotic'', or phantom matter, at least in a neighborhood 
  of a \wh\ throat (its narrowest place) has been formally established in a number 
  of theorems \cite{thorne,vis-book,HV97,fried,ws_book}, at least in cases  
  where the throat is a compact 2D surface of finite area \cite{HV97}.

  In theories generalizing GR, a number of examples of \wh\ solutions 
  without phantom matter are known, e.g., in the Einstein-Cartan theory 
  \cite{BGal15,BGal16},  in Einstein-Gauss-Bonnet theory \cite{GBo}, 
  in multidimensional gravity including brane-world models \cite{BKim03, BS16}, 
  etc. In all such cases phantom matter that would be necessary in GR is replaced 
  by geometric or field contributions absent in GR.  
  However, if our interest is in scales, let us say, from meters to kiloparsecs
  (which can be called macroscopic), it makes sense to adhere to GR since it is this 
  theory that is most reliably confirmed on such scales while its extensions 
  should probably work at very large densities and/or curvatures. 
  It is therefore of great interest to know whether or not \wh\ models can be obtained 
  in GR without phantom matter. And in addition, to be observed from distant 
  weakly curved regions of space, such solutions should be \asflat.
  
  Are there known phantom-free \wh\ solutions in GR? The answer is yes: 
  there are axially symmetric vacuum solutions such as the Kerr metric with 
  large angular momenta and Zipoy's static solution \cite{zipoy} as well as
  their generalizations with electromagnetic and scalar fields \cite{br-fab97, matos15}. 
  Being \asflat, they, however, contain a ring singularity around a disk that plays 
  the role of a throat, and it appears to be a shortcoming hardly better than the
  presence of phantom matter. Refs. \refcite{canf1,canf2} describe regular 
   \whs\  with a source in the form of a nonlinear sigma model, but they are
   asymptotically NUT-AdS instead of flatness. The \wh\ model described in 
  Ref. \refcite{schhein} does not contain phantoms but possesses closed timelike 
  curves and singularities.

  The above unpleasant features may be interpreted as manifestations of the so-called
  topological censorship theorems that forbid unusual topologies assuming that the
  WEC is observed and that the space-time is \asflat \cite{fried,woolgar}.  
  One of the ways to avoid this censorship is to consider configurations like 
  cosmic strings, infinitely stretched in some direction, and the simplest one among 
  them are those with \cy\ symmetry. In the longitudinal ($z$) direction there is no 
  flat infinity, unless the space-time is completely flat.  
  Static and stationary \cyl\ \wh\ solutions were discussed, 
  among many other papers, in Refs. 22--26
  (see also references therein). Quite a number of examples of phantom-free \cy\ \wh\
  solutions in GR have been found.

  However, an undesirable feature of almost all \cyl\ solutions is their nonflat asymptotic
  behavior. Even the Levi-Civita static vacuum solution is only \asflat\ if it is flat, to
  say nothing on more complex  solutions with matter. To solve this problem and to build 
  \asflat\ \wh\ models, it was suggested \cite{BLem13} to construct the model 
  space-time from three regions: a central region containing a throat and two 
  Minkowski regions (taken in proper coordinates) attached to this central \wh\
  solution through some surfaces (cylinders) $\Sigma_-$ and $\Sigma_+$. Such 
  surfaces inevitably possess some densities and pressures, and their stress-energy 
  tensors (SETs) are determined by jumps of the extrinsic curvature across them. 
  Exotic matter should then  be absent both in the internal region and on the surfaces 
  $\Sigma_-$ and $\Sigma_+$. A search for such models began in Refs. 23--26
  and is continued here. We will recall some old results and try to construct new models.
  
  The paper is organized as follows. Section 2 presents a general description of 
  stationary \cyl\ space-times, including the notion of \cy\ \whs, junction conditions 
  on possible discontinuities of the extrinsic curvature, and the WEC and NEC 
  applied to the corresponding surface SETs. In Section 3 we describe some particular 
  \wh\ solutions, try to match them to Minkowski regions on both sides of the throat,
  and test the fulfillment of the energy conditions. Section 4 is a conclusion.  

\section{Stationary \cy\ space-times}

\subsection{Basic relations}

  Consider the general stationary \cyl\ metric
\bearr                                                    \label{ds-rot}
         ds^2 = \e^{2\gamma(x)}[ dt - E(x)\e^{-2\gamma(x)}\, d\varphi ]^2
       - \e^{2\alpha(x)}dx^2 
	- \e^{2\mu(x)}dz^2 - \e^{2\beta(x)}d\varphi^2,
\ear
  where $x$, $z\in \R$ and $\varphi\in [0, 2\pi)$ are the radial, longitudinal and 
  angular coordinates, respectively. The radial coordinate $x$ admits reparametrization
  $x \to f(x)$, therefore its range depends on its choice (also called the ``gauge'') and 
  the geometry itself. The only off-diagonal component $E$  describes rotation 
  that induces a vortex gravitational field characterized by the angular velocity  
  $\omega(x)$ of a congruence of timelike world lines \cite{BLem13, kr2, kr4},
\beq                                                  \label{om}
          \omega = \half (E\e^{-2\gamma})' \e^{\gamma-\beta-\alpha},
\eeq
  for an arbitrary choice of the coordinate $x$ (the prime denotes
  $d/dx$). Furthermore, if the reference frame is comoving to matter 
  as it moves in the $\varphi$ direction, we the SET component
  $T^3_0$, is zero, hence (due to the Einstein equations) the Ricci tensor 
  component $R_0^3 \sim (\omega \e^{2\gamma+\mu})' = 0$, which results in 
\beq       	      					\label{omega}
	\omega = \omega_0 \e^{-\mu-2\gamma}, \cm \omega_0 = \const.
\eeq
  Then, according to \rf{om}, 
\beq                          \label{E}
	E(x) = 2\omega_0 \e^{2\gamma(x)} \int \e^{\alpha+\beta-\mu-3\gamma}dx,
\eeq
  and the Ricci tensor $(R\mN)$ has the nonzero components 
 \bear                 \label{Ric}
      R^0_0 \eql -\e^{-2\alpha}[\gamma'' + \gamma'(\sigma' -\alpha')] - 2\omega^2,
\nnv      
      R^1_1 \eql -\e^{-2\alpha}[\sigma'' + \sigma'{}^2 - 2U - \alpha'\sigma']+ 2\omega^2,
\nnv      
      R^2_2 \eql -\e^{-2\alpha}[\mu'' + \mu'(\sigma' -\alpha')],  
\nnv      
      R^3_3 \eql -\e^{-2\alpha}[\beta'' + \beta'(\sigma' -\alpha')] + 2\omega^2, 
\nnv
      R^0_3 = G^0_3 \eql   E \e^{-2\gamma}(R^3_3 - R^0_0), 
\ear
 with the following notations:
\beq
            \sigma = \beta + \gamma + \mu, \cm   
            U = \beta'\gamma'  + \beta'\mu' + \gamma' \mu'.
\eeq     
  The Einstein equations may be written in two equivalent forms 
\bearr    \label{EE1}
            G\mN \equiv R\mN - \Half \delta\mN R = -\kappa T\mN, \quad {\rm or}
\\   \lal                 \label{EE2}
            R\mN = - \kappa \tT\mN \equiv T\mN - \Half \delta\mN T.          
\ear    
  $\kappa = 8\pi G$ being the gravitational constant, $R$ the Ricci scalar, and $T$ 
  the trace of the SET. In what follows we mostly work with \rf{EE2}, but  it is also necessary 
  to use the constraint equation from \rf{EE1}, which is the first integral of the 
  other Einstein equations, containing only first-order derivatives of the metric functions:
\beq
	      G^1_1 = \e^{-2\alpha} U + \omega^2 = - \kappa T^1_1.
\eeq    
  
  As is clear from \rf{Ric}, the diagonal components of both Ricci ($R\mN$) and Einstein 
  ($G\mN$) tensors split into a sum of those for the static metric (i.e., (\ref{ds-rot}) with $E=0$) 
  and a contribution from $\omega$ \cite{BLem13}: 
\bearr 		\label{R-omega}
    		R\mN = {}_s R\mN + {}_\omega R\mN, \qquad
	{}_\omega R\mN = \omega^2 \diag (-2, 2, 0, 2),  \label{Ric-o}
\yyy     		\label{G-omega}  
		G\mN = {}_s G\mN + {}_\omega G\mN,  \qquad
   	{}_\omega G\mN = \omega^2 \diag (-3, 1, -1, 1),  \label{Ein-o}
\ear
  where ${}_s R\mN$ and ${}_s G\mN$ are the static parts. Each of the tensors 
  ${}_s G\mN$ and ${}_\omega G\mN$ satisfies the conservation law 
  $\nabla_\alpha G^\alpha_\mu =0$ written in terms of the static metric. It means that
  ${}_\omega G\mN/\kappa$ effectively acts as one more SET having exotic properties 
  (for instance, the effective energy density is $ -3\omega^2/\kappa <  0$), which is 
  favorable for the existence of wormhole solutions, as has been confirmed by 
  a few examples in Refs. \refcite{BLem13,BK15,kr4}.
  
  Notably, when solving the Einstein equations, it is sufficient to consider their 
  diagonal components; and the only nontrivial off-diagonal component ${0 \choose 3}$ 
  then holds automatically \cite{BLem13}.  

\subsection {Wormhole space-times}

  The metric \rf{ds-rot} is said to describe a \wh\ if either (i) the
  circular radius $r(x) = \e^{\beta(x)}$ has a regular minimum 
  (the corresponding surface $x=\const$ is called an $r$-throat) 
  and is large or infinite far from this minimum or (ii) the 
  same is true for the area function $a(x) = \e^{\mu+\beta}$ 
  (its minimum determines an $a$-throat) \cite{BLem09, BLem13}. 
  It is clear that if a \wh\  is \asflat\ on both extremes of the $x$ range,
  it contains both $r$- and $a$-throats.
  On the other hand, it has been shown \cite{BLem09, BLem13} that 
  in the static case ($E=0$) such a twice \asflat\ \wh\ should necessarily 
  contain a region where the density of matter is negative. In the general
  stationary case  ($E\ne 0$) it is not necessary, and in what follows we 
  will assume  $E \ne 0$.

  Let us, for certainty, use the \wh\ definition connected with the circular radius
  $r(x)$, which looks more evident: moving to smaller $r$, one approaches a 
  would-be axis of symmetry $r=0$, but, instead of reaching the axis, meets 
  a minimum of $r$ and its subsequent growth.
  
   Asymptotic flatness requires finite limiting values of  $\gamma(x)$ and $\mu(x)$ 
  combined with $\omega =0$, which cannot be achieved according to \rf{omega}.

  To obtain \asflat\ models, the following method  was suggested \cite{BLem13}:
  assuming that we have obtained a \cy\ \wh\ solution with a non-phantom matter 
  source $T\mN$, cut it on some regular cylinders $\Sigma_+\,(x=x_+)$ and 
  $\Sigma_-\,(x=x_-)$ on both sides of the throat and join it there to flat-space 
  regions extending to infinity. Due to discontinuities of the extrinsic
  curvature on such junction surfaces, they comprise thin shells with certain surface 
  SETs, and it remains to check whether these SETs satisfy the WEC and NEC.

\subsection{Junction surfaces and energy conditions}     

  Consider a surface $\Sigma$ ($x = x_s$) separating two regions  $\D_-$ ($x \leq x_s$)
  and $\D_+$ ($x \geq x_s$) with two different metrics of the form \rf{ds-rot}. 
  Since the metric on $\Sigma$ must be the same, be it calculated from $\D_-$ or $\D_+$,
  we have the following matching conditions:
\beq                                                       \label{ju-1}
      [\beta] = 0, \quad [\mu] = 0, \quad     [\gamma] = 0, \quad [E] =0,
\eeq
  with the usual notation for discontinuities: for any $f(x)$, $[f] = f(x_s+0) - f(x_s -0)$. 
  The conditions \rf{ju-1} allow us to identify the coordinates $t, z, \phi$ in the whole space.
  Meanwhile, the choice of radial coordinates in $\D_+$ and $\D_-$ may be different, 
  but it is unimportant since the quantities involved in all matching conditions are
  insensitive to this choice.

  The next step is to determine the material content of the surface $\Sigma$ using
  the Darmois-Israel formalism  \cite{israel-67,BKT-87}: in our case of a 
  timelike surface, the SET $S_a^b$ is expressed in terms of the extrinsic curvature 
  $K_a^b$ as
\bearr                                                        \label{ju-2}
	S_a^b =  (8\pi G)^{-1} [\tK_a^b], \quad
                  		\tK_a^b := K_a^b - \delta_a^b K^c_c, 		
\ear
  where $a, b, c = 0, 2, 3$. The question is whether the surface SET on $\Sigma$ 
  satisfies the WEC whose requirements are
\beq                                                             \label{WEC}
	S_{00}/g_{00} = \sigma_s \geq 0, \qquad\
				S_{ab}\xi^a \xi^b \geq 0,
\eeq
  where $\sigma_s$ is the surface energy density and $\xi^a$ an arbitrary null
  vector on $\Sigma$. The second inequality makes the content of the NEC as a 
  part of the WEC, and taken together, they provide $\sigma_s \geq 0$ in any reference 
  frame on $\Sigma$.
  
  It is straightforward to find the following nonzero components of 
  $K_{ab} = \half \e^{-\alpha} g'_{ab}$ on a surface $x=x_s$:
\bearr                             \label{Kab}
      K_{00} = \e^{-\alpha+2\gamma} \gamma',    
\nnn      
      K_{03} = -\half \e^{-\alpha} E',
\nnn
	K_{22} = -\e^{-\alpha + 2\mu} \mu',
\nnn 	
	K_{33} = -\e^{-\alpha+2\beta}\beta' + \e^{-\alpha-2\gamma}(EE'- E^2\gamma').
\ear      
   Since $K = g^{ab} K_{ab} = \e^{-\alpha}(\beta'+\gamma'+\mu')$, it 
   is easy to obtain the components of $\tK_{ab} = K_{ab} - g_{ab}K$, to be further
   used instead of $S_{ab} = [\tK_{ab}]/\kappa$ for WEC verification:
\bearr                              \label{tKab}
      \tK_{00} = - \e^{-\alpha+2\gamma} (\beta' + \mu'),    
\nnn      
      \tK_{03} = -\half \e^{-\alpha} E' + E\e^{-\alpha}(\beta'+\gamma'+\mu'),
\nnn
      \tK_{22} = \e^{-\alpha + 2\mu} (\beta'+\gamma'),
\nnn 
      \tK_{33} = \e^{-\alpha+2\beta}(\gamma'+\mu') 
	+ \e^{-\alpha-2\gamma}[EE'- E^2(\beta'+2\gamma'+\mu')].. 	      
\ear      
  From \rf{tKab} we obtain the condition $\sigma_s \geq 0$ in the form
\beq                                \label{dens}  
 		 [\e^{-\alpha} (\beta' + \mu')] \leq 0.
\eeq 		 
  Meanwhile, the NEC validity must be proved for {\it any\/} null vector $\xi^a$ on $\Sigma$. 
  To cover all null directions on $\Sigma_\pm$, we should take a family of vectors $\xi^a$
  depending on one parameter, say, $h$, and to try to choose such parameters in our 
  metrics that the conditions \rf{WEC} will hold  for {\it any\/} $h$ from the appropriate range. 
  This proves to lead to rather bulky calculations.   
  
  However, the NEC fulfillment may be verified in another way: if in the 
  comoving reference frame the density $\sigma_s$ and the pressures $p_i$ in 
  mutually orthogonal directions satisfy the inequalities 
\beq   				\label{WEC3}
  	\sigma_s \geq 0, \cm  \sigma_s + p_i \geq 0, 
 \eeq 	
  then the WEC holds. A difficulty is that $\Sigma$ is in general not described in 
  a comoving frame. 
  
  It is still not necessary to find an explicit transformation to the comoving frame for matter 
  residing on $\Sigma$, which can be an uneasy task. Instead, it is sufficient to find 
  the values of $\sigma_s$ and the pressures $p_z$, $p_\varphi$ in such a frame as 
  eigenvalues of the surface SET and then check whether or not \rf{WEC} holds.
  These eigenvalues should be calculated in a local Minkowski (tangent) space,
  to avoid distortions from the nontrivial metric. 

  Let us find the tangent-space (triad) components of $\tK_{ab}$ using 
  the following orthonormal triad on $\Sigma$:
\beq  				\label{triad}
              e_{(0)}^a = (\e^{-\gamma},0,0);    \qquad
              e_{(2)}^a = (0, \e^{-\mu}, 0); \qquad
              e_{(3)}^a = (E\e^{-\beta-2\gamma}, 0, \e^{-\beta})           
\eeq      
   (the parentheses mark triad indices).
   The triad components $\tK_{(mn)} = e_{(m)}^a e_{(n)}^b \tK_{ab}$ turn out to 
   be surprisingly simple and may be represented by the matrix
\beq  				\label{tK-matrix}                          
	     (\tK_{(mn)}) = \begin{pmatrix}
	     					-\e^{-\alpha}(\beta'+\mu')  &  0  & -\omega \\
	     					 0              &      \e^{-\alpha}(\beta' + \gamma') & 0 \\
	     					 -\omega   &   0     & \e^{-\alpha}(\gamma'+ \mu')	     					     \end{pmatrix}.		     
\eeq      
   The shell matter SET consists of discontinuities of these matrix elements divided
   by $\kappa$. The matrix of these discontinuities has the same structure as \rf{tK-matrix}: 
\beq   				\label{matrix}      
		([\tK_{(mn)}]) =                   
		\begin{pmatrix}            \ a\ \ & 0\ \ & d \ \\
						      \ 0\ \ & b\ \ & 0 \ \\
						      \ d\ \ & 0\ \ & c \ \end{pmatrix},
\eeq 
   and its eigenvalues are easily found as roots of its characteristic equation:
\beq   				\label{roots}                          
             \Big(\Half(a+c+\sqrt{(a-c)^2 +4 d^2}),\  b,\ \Half(a+c -\sqrt{(a-c)^2 +4 d^2}) \Big),
\eeq      
   The SET in question has the form $S_{(mn)} = \diag (\sigma_s, p_z, p_\varphi)$, 
   but there is a small problem of which eigenvalues \rf{roots} should be identified with
   particular SET components. To fix it, we notice that in the absence of rotation,
   that is, if $d=0$, we have $(\sigma_s, p_z, p_\varphi) \propto (a,b,c)$.
   Accordingly, we take
\beq         \label{roots1}
	(\sigma_s, p_z, p_\varphi) \propto 	
	\Big(\Half(a+c+\sqrt{(a-c)^2 +4 d^2}),\  b,\ \Half(a+c -\sqrt{(a-c)^2 +4 d^2}) \Big),
\eeq
  under the assumption $a-c >0$. (Otherwise we must interchange $a$ and $c$.)
  Then the WEC requirements read
\bearr        \label{Wa} 
	a+c+\sqrt{(a-c)^2 +4 d^2} \geq 0,
\yyy          \label{Wb}    
	a+c+\sqrt{(a-c)^2 +4 d^2} + 2b \geq 0,
\yyy            \label{Wc}
	a + c \geq 0.          
\ear                                 
   One can notice that $a$ coincides with the quantity $\kappa\sigma_s$ \rf{dens} 
   calculated in the initial noncomoving frame, hence we must also have $a\geq 0$.

\subsection{Minkowski regions}

  We are going to use Minkowski space-time regions $\M_{\pm}$ around ``internal''
  \wh\ regions. The Minkowski  metric should be taken in a rotating reference frame to 
  enable matching with \rf{ds-rot} where $E\ne 0$. It can be obtained from 
  the inertial-frame metric $ds_{\rm M}^2 = dt^2 - dX^2 - dz^2 - X^2 d\varphi^2$  
  by substituting $\varphi \to \varphi + \Omega t$, where $\Omega = \const$ is the angular 
  velocity:
\beq                                                          \label{ds_M}
	      ds_{\rm M}^2 = dt^2 - dX^2 - dz^2 - X^2 (d\varphi + \Omega dt)^2.
\eeq
  The relevant quantities in the notations of (\ref{ds-rot}) are
\bearr                                                   \label{M-param}
      \e^\alpha = 1, \qquad   \e^{2\gamma} =  1 - \Omega^2 X^2,\qquad
      \e^{2\beta} = \frac{X^2}{1 - \Omega^2 X^2},
\nnn
      E = \Omega X^2, \qquad      \omega = \frac{\Omega}{1 - \Omega^2 X^2}.
\nnn 
      \gamma' = -\frac{\Omega^ X}{1 - \Omega^2 X^2}, \qquad   \mu' =0, \qquad
      \beta'  = \frac 1X  + \frac{\Omega^2 X}{1 - \Omega^2 X^2}.
\ear
  This metric is stationary and can be matched to an internal metric at $|X| <  1/|\Omega|$, 
  such that the linear rotational velocity is smaller than $c$.

  A good exercise is to verify that direct matching of two Minkowski regions with 
  the metric \rf{ds_M} leads to a thin-shell \wh\ with negative density of on-shell matter. 
  We identify the surface $X = X_0 > 0$ that bounds the region $\M_+ : x \geq X_0$
  with the surface $X = -X_0 < 0$ that bounds the region $\M_- : x \leq -X_0$ and obtain
\beq
	   - \kappa\sigma_s = [\e^{-\alpha} (\beta' + \mu')] = [\beta'] 
	   			= \frac{2}{|X_0|(1 - \Omega^2 X_0^2)} >0,
\eeq    
  that is, $\sigma_s < 0$.
  
\section {Some particular models}

\subsection{Models with scalar fields}

  Minimally coupled scalar fields are described by the Lagrangian
\beq
	L_s = 2\eps g\MN \d_\mu\phi \d_\nu\phi  -2 V(\phi)
\eeq
  where $V(\phi)$ is an arbitrary potential, and $\eps = \pm 1$ distinguishes
  normal (canonical) scalar fields ($\eps = +1$) and phantom ones ($\eps=-1$).

  In the geometry \rf{ds-rot}, assuming $\phi = \phi(u)$,
  the scalar field SET has the form
\beq                                                          \label{SET-phi}
     T\mN (\phi) = \eps\e^{-2\alpha}\phi'{}^2 \diag(1,\ -1,\ 1,\ 1) + \delta\mN V(\phi).
\eeq
  There are many exact solutions to the Einstein-scalar field equations with 
  and without a potential, including \wh\ ones \cite{BLem13, BK15},
  but it turns out that the construction with two external Minkowski regions 
  around a \wh\ region does not lead to phantom-free models. The corresponding
  no-go theorem \cite{kb-conf16} is valid for any SETs having the property
  $T^0_0 = T^3_3$, which is true for scalar fields with any $V(\phi)$ and $\eps$. 
  The theorem is proved using the equation $R^0_0 = R^3_3$. Choosing the 
  harmonic radial coordinate in the metric \rf{ds-rot}, specified by the condition
\beq                        \label{harm}
		\alpha = \beta + \gamma + \mu,
\eeq      
  we obtain the equation $R^0_0 = R^3_3$ in the Liouville form 
\bear                                                         \label{b-g}
       \beta'' - \gamma'' = 4\omega_0^2 \e^{2\beta - 2\gamma},
\ear
  and its solution can be written as
\bearr 								\label{s1}
	\e^{\eta} := \e^{\gamma-\beta} = 2\omega_0 s(k,x),  
\nnn
	s(k,x) :=
	\vars{
	k^{-1} \sinh kx, & \  k > 0,\ \ x \in \R_+;\\
		      x, & \ k=0, \ \ x \in \R_+; \\
	 k^{-1} \sin kx, & \ k<0, \ \ 0 < x < \pi/|k|,
	}
\ear       
   A further analysis of WEC requirements shows that  \cite{kb-conf16}
   one of them can hold on $\Sigma_+$ only if $k >0$ in  the internal 
   solution, while on  $\Sigma_-$ it definitely implies $k <0$ (or vice versa). 
   This means that whatever particular solution (with fixed parameters 
   including $k$) is taken to describe the internal region, the WEC 
   requirements cannot hold simultaneously on $\Sigma_+$ and  $\Sigma_-$.

\subsection{Perfect fluids with $p = w\rho$}    
            
   Let us now consider another kind of matter for the internal \wh\ region, an 
   isotropic perfect fluid with the equation of state $p = w\rho$, $w=\const$, so that
\bearr                        \label{SET-w}
		T\mN = \rho \diag (1, -w, -w, -w), 
\nnn
		\tT\mN = T\mN - \Half \delta\mN T^\sigma_\sigma
		           =  \frac \rho 2 \diag(3w+1, w-1, w-1, w-1),
\ear            
   plus the off-diagonal component $T^0_3 =  E \e^{-2\gamma}(T^3_3 - T^0_0)$.
   The conservation law $\nabla_\nu T\mN =0$ has the form
\beq                         \label{cons-w}
                p' + (\rho + p) \gamma' = 0  \ \then \  w \rho' + (1+w)\gamma' =0. 
\eeq   
   (the same as in the static case). For $w \ne 0$ it gives
\beq                          \label{rho-w}
	         \rho = \rho_0 \e^{-\gamma(w+1)/w}, \cm \rho_0 = \const,
\eeq            
    and, in terms of the harmonic coordinate $x$ \rf{harm}, the Einstein equations 
    for the metric \rf{ds-rot} can be written as
\bear           \label{R00-w}
  	\e^{-2\alpha} \gamma'' + 2\omega^2 \eql \frac {3w+1}{2}\kappa\rho ,
\\ 	\label{R22-w}
	\e^{-2\alpha} \mu'' \eql \frac {w-1}{2}\kappa\rho,
\\ 	\label{R33-w}
	\e^{-2\alpha} \beta'' - 2\omega^2 \eql \frac{w-1}{2} \kappa\rho,
\\ 		\label{G11-w}
	\e^{-2\alpha} (\beta'\gamma' + \beta'\mu' + \gamma'\mu') 
	                + \omega^2 \eql  w\kappa\rho.
\ear
   Also recall that $\omega = \omega_0 \e^{-\mu-2\gamma}$.
   
   There are exact solutions with the metric \rf{ds-rot} and different perfect 
   fluids \cite{kb79, santos82, skla, iva02a}, but their general discussion is out 
   of the scope of this paper. Let us only make some remarks on the
   simplest cases.
   
   First, in the case of dust ($w=0$) the solution is rather easily obtained
   since the conservation law \rf{cons-w} in this case leads to $\gamma' =0$,
   hence $\gamma =0$ without loss of generality. Next, a sum of 
   \rf{R00-w} and \rf{R33-w} yields $\beta'' + \gamma''=0$, whence 
   $\beta'' = 0$, and $\beta$ is a linear function that cannot have a minimum,
   so \wh\ solutions cannot be expected. Studies of rotating dust solutions 
   can be found in Refs. \refcite{iva02b,exact-book,grif}, containing further bibliography. 
   
   Second, the case $w = -1$ corresponds to a cosmological constant, for 
   which solutions are well known and well studied 
   \cite{exact-book, lanczos, lewis, mac98}. For our discussion it is important 
   that \wh\ solutions do exist in this case \cite{BLem13},
   but since, just as for scalar fields, here again $T^0_0 = T^3_3$, such
   solutions do not lead to phantom-free models.
   
 \subsection{A model with stiff matter, $p = \rho$}   
 
   The special case $w=1$ of a perfect fluid corresponds to maximally stiff matter 
   compatible with causality, in which the velocities of sound and light coincide. 
   This circumstance makes easier a search for exact solutions, including inhomogeneous 
   and wave ones \cite{kb-stiff}; one can also mention various applications of stiff
   matter in theoretical cosmology, see, e.g., the recent papers Refs. \refcite{stiff1,stiff2}
   and references therein.
   
   Returning to our perfect fluid equations, at $w=1$ from \rf{R22-w} we have 
\beq   
	\mu'' =0  \ \then \                  \mu = mx + \mu_0, \cm m, \mu_0=\const,
 \eeq  
   and we put $\mu_0=0$ by rescaling the $z$ axis. Then from \eqs \rf{R00-w} and 
   \rf{R33-w} we obtain
\beq     
                   \beta'' = 2\omega_0^2 \e^{2\beta-2\gamma}, \cm
                   \beta'' + \gamma'' = 2\kappa\rho_0 \e^{2\beta + 2mx}.  
\eeq          
  It is not easy to find a general solution to these equations, but a special solution of interest 
  is found by assuming $m=0$ (enabling symmetry under reflections $x \to -x$)
  and $\gamma \equiv 0$. It then follows
\beq                \label{be''}
		  \beta'' = 2\omega_0^2 \e^{2\beta},    \cm \kappa\rho_0 = \omega_0^2.
\eeq      
   This Liouville equation has three branches of solutions (see \rf{s1}), but only in one of 
   them the function $\beta(x)$ has a minimum and is of interest in search for \wh\ models.  
   This solution can be written as
\beq
		\e^{\beta} = \frac {k}{\sqrt{2\omega_0^2}\cos (kx)}, \cm k = \const> 0,
		\cm x \in (-\pi/2, \pi/2),  
\eeq      
   where we suppress another integration constant by choosing the zero point of $x$. 
   It remains to find $E$ using the expression \rf{E}:
\beq           \label{E1}
		E = \frac{1}{\omega_0} \int \frac{k^2 dx}{\cos^2(kx)} = \frac{k}{\omega_0} \tan (kx),
\eeq
  where the integration constant is chosen so that $E(x)$ is an odd function for convenient
  matching with the exterior metric. Thus the metric is known completely.
  
  It is convenient to pass on to another radial variable, $y = k \tan(kx)$, so that 
\beq                      \label{x-y}
             dx = \frac{dy}{k^2 + y^2}, \cm	\e^{2\beta} = \frac{k^2+y^2}{2\omega_0^2 }, 
             \cm E = \frac{y}{\omega_0},
\eeq
  and the metric takes the form
\beq                                       \label{ds1}  
  	   ds^2 = \bigg(dt -  \frac{y}{\omega_0} d\varphi\bigg)^2 
  	   - \frac{dy^2}{2\omega_0^2(k^2+y^2)} - dz^2 - (k^2 + y^2) \frac{d\varphi^2}{2\omega_0^2}.
\eeq  
    This solution is regular in the whole range $y\in \R$, but at $y^2 > k^2$ we have 
    $g_{33} > 0$, hence the coordinate circles parametrized by $\varphi$ are timelike and 
    violate causality.
  
    We are now almost ready to match this metric to \rf{ds_M} at some surface $y=y_0$ 
    identified with some $X=X_0$ in \rf{ds_M} according to the conditions \rf{ju-1}. 
    However, to obtain $[\gamma]=0$ it is necessary to change the time scale in
    \rf{ds1}, such that 
\beq                 \label{t-tau}
              dt = \sqrt{P} d\tau, \cm P = \const < 1
\eeq        
    Then we can identify the time $\tau$ on the surfaces
    $\Sigma_\pm$ with $t$ from the external metric and provide $[\gamma]=0$. For the 
    internal metric \rf{ds1} we then have, instead of the last relation in  \rf{x-y},
    $E = \sqrt{P} y/\omega_0$. (One can verify that this new $E$ is the off-diagonal 
    component $-g_{\tau \phi}$.) 
    
    Consider for certainty the surface $\Sigma_+$ ($y=y_0 >0$, $X=X_0 > 0$).
    The conditions $[\gamma]=0$, $[E] =0$ and $[\beta] =0$ give, respectively,
\beq                \label{match1}
                    P = 1 - \Omega^2 X^2, \cm
                    \Omega X^2 = \frac{y\sqrt{P}}{\omega_0}, \cm 
                    \frac{k^2+y^2}{2\omega_0^2} = \frac{X^2}{P}.
\eeq       
     where we assume $\omega_0 > 0$ and, without risk of confusion, omit the index 
     ``zero'' at $X$ and $y$.  We notice that there are four independent parameters 
     ($P,  k, \omega_0, y$) in the internal  metric \rf{ds1} after the substitution \rf{t-tau}
     and two parameters ($\Omega, X$) 
    in the external metric \rf{ds_M}. These six parameters are connected by three equalities
    \rf{match1}. Let us choose the following three parameters as independent ones: 
    $X = X_0$, having the dimension of length, as a length scale, $y =y_0$, and $P$. 
    We then obtain:
\beq            \label{param1}  
		  \Omega = \frac{\sqrt{1-P}}{X} , \cm 
		  \omega_0 = \frac{\sqrt{P} y}{\sqrt{1-P}X}, \cm   
		  k^2 = y^2 \frac{1+P}{1-P},
\eeq        
  and, as a result, the quantities $a,b,c,d$ from \rf{matrix}--\rf{Wa} are expressed as 
\bearr                \label{abcd1}
	a = \frac{P^{3/2}y  -1}{PX}, \quad\
	b = \frac{1 - y\sqrt{P}}{X}, \quad\
	c = \frac{P-1}{PX}, \quad\
	d = \frac{\sqrt{P}y - 1 + P}{XP\sqrt{1-P}}.  
\ear      
  The factor $1/X$ is common and does not affect any inequalities of interest. 
  The condition $a >0$ implies $y > P^{-3/2}$. However, since $c <0$, the 
  requirement \rf{Wc} $a+c \geq 0$ gives a stronger restriction 
\beq            \label{y_1} 
		y \geq \frac{2 - P}{P^{3/2}. }
\eeq    
  Then  \rf{Wa} manifestly holds, while \rf{Wb} requires verification due to $b < 0$,
  An inspection shows that \rf{Wb} does hold owing to \rf{y_1}.
  
  Thus under the condition \rf{y_1} the WEC is fulfilled on $\Sigma_+$. What changes on 
  $\Sigma_-$ specified by $X = -X_0 < 0$ and $y= -y_0 <0$?
  In all parameters  \rf{abcd1} the factor $1/X$ becomes negative. But 
  simultaneously changes the sign of all discontinuities: while on $\Sigma_+$ we took
  $[f] = f_{\rm out} - f_{\rm in}$ for any quantity $f$, on $\Sigma_-$ we must take the 
  opposite. Therefore, the parameters $a,b,c$ have the same form as in \rf{abcd1} 
  but with $X$ replaced by $|X|$ (recall that we denote, as before, $y = y_0 >0$).
  For $d = -[\omega]$ we must take into account that, according to \rf{match1},
  $\Omega (\Sigma_-) = -\Omega (\Sigma_+)$, while in the internal solution 
  $\omega (\Sigma_-) = \omega (\Sigma_+)$, therefore, on $\Sigma_-$
\[
	d \ \mapsto \ d_- = - \frac {1 - P + |y| \sqrt{P}}{|X| P \sqrt{1-P}},
\]    
  so that $|d_-| > |d|$, which makes it even easier to satisfy the requirement \rf{Wb}. 
  As a result, all WEC requirements are satisfied under the same condition \rf{y_1},
  and we obtain a completely  phantom-free \wh\ model.
  
  It is also important that by \rf{param1} we have $y_0^2 < k^2$, hence $y^2 < k^2$
  in the whole internal region, and there are no closed timelike curves.
  
\subsection {A model with an anisotropic fluid}    
    
  Let us briefly describe one more model, obtained in Ref.\,\refcite{BK18} with a source
  in the form of an anisotropic fluid having the SET 
\beq  \label{SET-2}
	   T\mN = \rho \diag (1, -1, 1, -1)  \quad \oplus \quad 	   
	   T^0_3 = -2\rho E \e^{-2\gamma},
\eeq	   
  chosen by analogy with that of a $z$-directed  magnetic field in static 
  cylindrical symmetry (which cannot be directly extended to the metric \rf{ds-rot} 
  with $E\ne 0$). From the conservation law $\nabla_\mu T^\mu_1 =0$ it 
  follows\footnote
  		{A full description of the anisotropic fluid formalism in the metric
  		  \rf{ds-rot} can be found, e.g., in Ref. \refcite{santos}.}  
\beq        \label{rho-2}
       \rho = \rho_0 \e^{-2\gamma - 2\mu},  \quad\ \rho_0 = \const > 0,
\eeq 
  The Einstein equations are solved \cite{BK18} using the harmonic radial coordinate
  \rf{harm}, and the solution has the form
\bearr                                       \label{sol-2}
      r^2 \equiv \e^{2\beta}= \frac{r_0^2}{Q^2 (x_0^2-x^2)}, \cm
		\e^{2\gamma} = Q^2 (x_0^2-x^2), 
\nnnv 
			\e^{2\mu} = \e^{2mx} (x_0 -x)^{1-x/x_0} (x_0 +x)^{1+ x/x_0},	
\nnnv			
	E = \frac {r_0} {2x_0^2} \bigg[ 2x_0 x 	+ (x_0^2 - x^2) \ln\frac{x_0+x}{x_0 -x} \bigg],
\nnn	 \cm		
	x_0 := \frac{|\omega_0|}{\kappa\rho_0 r_0}, \qquad\ Q^2 := \kappa \rho_0 r_0^2,      
\ear
  The introduced constants $x_0$ and $Q$ are dimensionless. 
  The solution contains integration constants $\omega_0$, $\rho_0$, $r_0$ and $m$, 
  the coordinate $x$ ranges from $-x_0$ to $x_0$. The circular radius 
  $r \to \infty$ as $x\to \pm x_0$, thus confirming a \wh\ nature of the geometry, 
  but $x = \pm x_0$ are curvature singularities, where the Kretschmann invariant
  behaves as $|x_0-x|^{-4}$.  
  
  To construct an \asflat\ configuration, we assume $m=0$, making the solution 
  symmetric with respect to the throat $x=0$. Its matching at some $x= \pm x_s < x_0$ 
  to $\M_+$ and $\M_-$ at $X = \pm X_s$  leads to the relations 
\bearr             \label{jun-2}
		X = r_0, \cm  
		Q^2 (x_0^2 -x^2) = 1 - \Omega^2 X^2 =: P, 
\nnn
		2x_0^2 \sqrt{1-P} = 2xx_0 + (x_0^2 -x^2) \ln \frac {x_0+x}{x_0-x},		
\ear    
  where we omit the index $s$ near $x$ without risk of confusion. The condition $[\mu] =0$
  makes us change the $z$ scale in $\M_\pm$ so that there $-g_{zz} = M^2 := \e^{2\mu(x_s)}$
  taken from the internal metric.

  It is convenient to use the ratio $y = x_s/x_0$ and to introduce the notation 
  $L(y) = \ln[(1+y)/(1-y)]$. Then we have   
\bearr
	 M = M(y) = \big(1-y\big)^{-(1-y)/2}\big(1+y\big)^{-(1+y)/2},
\nnnv                 
	P =P(y)  = (1-y^2)\Big[1 - y L(y) - \frac 14 (1-y^2) L^2(y)\Big].
\ear  
   Using the conditions \rf{jun-2} and these notations, we can write the quantities 
   $a,b,c,d$ used in the WEC requirements \rf{Wa}--\rf{Wc} on $\Sigma_\pm$ as follows:
\bearr
	a = [-\e^{-\alpha}(\beta'+\mu')]  = 
	  - \frac 1{P(y)} + \frac {M(y)}{x_0^2}\bigg(\frac{y}{1-y^2} + \Half L(y)\bigg),
\\ \lal
	b = [\e^{-\alpha}(\beta' + \gamma')] = 1,
\\ \lal     
        c= [\e^{-\alpha}(\gamma'+ \mu')] 
        = - \frac 1{P(y)} + 1 + \frac {M(y)}{x_0^2}\bigg(\frac{y}{1-y^2} - \Half L(y)\bigg),
\\ \lal
	d =- [\omega] =  - \frac{\sqrt{1-P(y)}}{P(y)} \pm  \frac{M(y)}{x_0^2(1-y^2)}, 
\ear      
  where we have ignored the insignificant factor $1/r_0$ appearing in each of them.
  
  The expressions for $a,b,c,d$ depend on two parameters, $x_0$ and $y$, and  
  $a,b,c$ are the same on $\Sigma_+$ and $\Sigma_-$, while it turns out that the 
  value of $d$ does not affect the validity of the conditions \rf{Wa}--\rf{Wc}: 
  actually, if $a > 0$ and $a +c >0$, then \rf{Wb} holds. All quantities in these conditions are dimensionless, and it can be found  that \cite{BK18}

\medskip  
  \noi$\bullet$ The condition $0< P(y) < 1$, required by construction, holds for $0< y < 0.564$
   (all numerical estimates are approximate);
  
\medskip    
  \noi$\bullet$ The conditions $a > 0$ and $a >c$ hold in quite a large range of $x_0$ 
  and $y$, for example, for $x_0 = 0.5,\ y \in (0.15, 0.47)$ and for $x_0 = 0.3,\ y \in (0.05, 0.53)$.

\medskip    
  \noi$\bullet$ The condition $a+c >0$ also holds in almost the same range of the 
  parameters, e.g, for $x_0 = 0.5,\ y \in (0.15, 0.38)$ and for $x_0 = 0.3,\ y \in (0.05, 0.51)$.
  
  Thus there is a significant range in the parameter space ($x_0, y$) in which this
  \asflat\ \wh\ model completely satisfies the WEC. It can also be verified that $g_{33} <0$
  in the internal solution between $\Sigma_-$ and $\Sigma_+$, hence the model does not 
  contain closed timelike curves.

\section {Conclusion}

  Our consideration demonstrates the possible existence of twice asymptotically flat
  cylindrically symmetric wormholes without exotic matter and without closed timelike curves
  in general relativity. In stationary cylindrically symmetric space-times it appears possible 
  to obtain a number of exact wormhole-type solutions to the Einstein equations with 
  various material sources (isotropic and anisotropic fluids, scalar and electromagnetic
  fields, etc.). However, to provide asymptotic flatness is a separate problem, and here,  
  following  Ref. \refcite{BLem13}, we construct the whole space-time as a union of 
  three regions: a central region with a throat and two Minkowski regions (taken in  
  appropriate rotating reference frames), attached to the central region using
  the Darmois-Israel formalism with the corresponding junction conditions.

  An important task is to find such models where the effective SET on the junction surfaces
  is not exotic, i.e., does not violate the WEC/NEC. We analyze this property by finding
  eigenvalues of the extrinsic curvature tensor discontinuities that represent the shell matter 
  SET in its comoving reference frame.
  
  Our analysis shows the existence of non-empty sets of parameters for which the 
  WEC/NEC are indeed respected. Also, the negative sign of $g_{\phi\phi}$ proves 
  the absence of closed timelike curves in the whole space-time under consideration.
  Moreover, it seems that our models do not violate the topological censorship theorems
  that restrict the existence of nontrivial space-time topologies assuming that the 
  WEC holds true  and the space-time is asymptotically flat. A possible reason is that 
  our cosmic string-like models are not asymptotically flat in the longitudinal direction
  and are thus not completely \asflat.
  
  An attractive opportunity is that such \asflat\ configurations, by analogy with cosmic 
  strings, could form  loops with a size much larger that their inherent  characteristic length
  parameters, so that cylindrical symmetry could be approximately valid. 
  
   In our view, it is of interest to seek new potentially realistic sources for such geometries 
  and to further study their mathematical and physical properties including possible 
  observational effects.

\subsection*{Acknowledgments}

  This publication was supported by the RUDN University program 5-100. 
  The work of KB was also performed within the framework of the Center FRPP 
  supported by MEPhI Academic Excellence Project (contract No. 02.a03.21.0005,
  27.08.2013).

\end{document}